\documentclass[aps,prl,twocolumn,superscriptaddress]{revtex4}
\usepackage{amsmath}
\usepackage{graphicx}
\usepackage{fancybox}
\usepackage{subfigure}

\DeclareMathOperator{\erf}{erf}
\DeclareMathOperator{\erfc}{erfc}

\begin{document}


\title{Attraction Between
  Like-Charged Walls:\\ Short-Ranged Simulations Using Local Molecular Field Theory}


\author{Jocelyn M. Rodgers}
\affiliation{Institute for Physical Science and Technology, University of
  Maryland, College Park, Maryland 20742}
\affiliation{Chemical Physics Program, University of Maryland, College Park, Maryland 20742}

\author{Charanbir Kaur}
\altaffiliation[Present address: ]{McCombs School of Business, University of Texas at Austin,
  Austin, TX 78712}
\affiliation{Institute for Physical Science and Technology, University of
  Maryland, College Park, Maryland 20742}

\author{Yng-Gwei Chen}
\altaffiliation[Present address: ]{Laboratory of Chemical Physics and National
Institute of Diabetes and Digestive and Kidney Diseases, Bldg. 5, National
Institutes of Health, Bethesda, MD 20892}
\affiliation{Institute for Physical Science and Technology, University of
  Maryland, College Park, Maryland 20742}
\affiliation{Department of Physics, University of Maryland, College Park, Maryland 20742}

\author{John D. Weeks}
\affiliation{Institute for Physical Science and Technology, University of
  Maryland, College Park, Maryland 20742}
\affiliation{Department of Chemistry and Biochemistry, University of Maryland,
  College Park, Maryland 20742}


\date{\today}

\begin{abstract}
  Effective attraction between like-charged walls mediated by counterions is
  studied using local molecular field (LMF) theory.  Monte Carlo simulations
  of the ``mimic system'' given by LMF theory, with short-ranged ``Coulomb core"
  interactions in an effective single particle potential incorporating a mean-field average of the
  long-ranged Coulomb interactions, provide a direct test of the theory, and
  are in excellent agreement with more complex simulations of the full Coulomb
  system by Moreira and Netz [Eur.~Phys.~J.~E {\bf 8}, 33 (2002)].  A simple,
  generally-applicable criterion to determine the consistency parameter
  $\sigma_{min}$ needed for accurate use of the LMF theory is presented.
\end{abstract}


\maketitle

\begin{figure}[tbp]
  \centering
  \subfigure[Full Model System]{
    \label{fig:FullModel}
    \includegraphics*[width=3.375in]{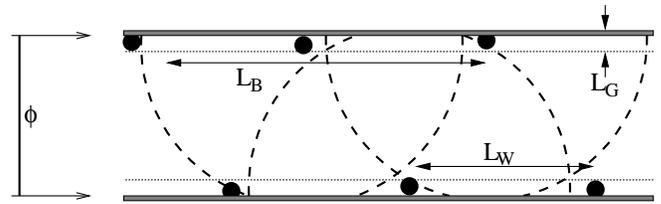}}
  \vspace{0.3cm}
  \subfigure[LMF Mimic System]{
    \label{fig:LMFModel}
    \includegraphics*[width=3.375in]{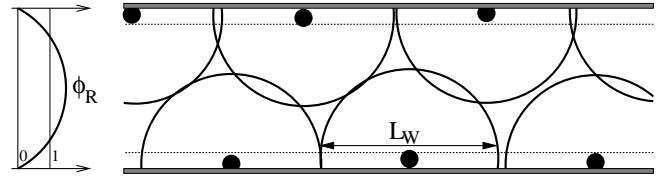}}
  \caption{Ion distributions and length scales for a moderate coupling case
    where the attractions are just beginning to develop ($\xi=20$ and $d=10$).
    The full model system shown in~\subref{fig:FullModel} consists of point
    counterions represented by points that neutralize two charged hard walls,
    with dashed circles for $L_B$ (only 3 $L_B$ circles are shown for clarity) and
    dotted lines for $L_G$.  The electrostatic potential $\phi$ due to the walls is 0. 
    The LMF mimic system shown in~\subref{fig:LMFModel} has Coulomb core
    interactions with a range of $\sigma_{min}$ proportional to the
    spacing $L_W$ (indicated by solid circles), and a modified
    wall potential $\phi_R$ that accounts for the remaining long-ranged
    interactions.}
  \label{fig:Model}
\end{figure}

Effective attractions between like-charged objects are quite common and have been extensively
studied~\cite{LevinReview,DNAElectrostatics,RouzinaBloomfield,NetzReview2005}.
For example, highly charged DNA is densely packed in cell nuclei via
positively-charged intermediaries, and, \emph{in vitro}, DNA may be condensed
into toroids by adding sufficient concentrations of divalent or trivalent
counterions. In this paper we use local molecular field (LMF) theory to study
one of the simplest models that exhibits like-charged attraction --- two
uniformly-charged walls with neutralizing point counterions, as shown in
Fig.~\ref{fig:FullModel} with length scales that will be discussed later.

LMF theory defines a general mapping that relates the structure and
thermodynamics of a nonuniform system with long-ranged intermolecular
interactions in an external field $\phi$ to those of a simpler ``mimic
system'' with short-ranged interactions in the presence of an effective field
$\phi_R$, as qualitatively shown in Fig.~\ref{fig:LMFModel}. $\phi_R$
accounts for the averaged effects of the long-ranged interactions and
self-consistently depends on the nonuniform density the field
induces~\cite{WeeksYBG2, WeeksLMF}.  This approach is
particularly useful for systems with Coulomb interactions, because one can
choose specific slowly-varying, long-ranged components of the Coulomb
interactions that are especially well-suited for the mean-field
average.  The remaining short-ranged
``Coulomb core" components combine with other existing short-ranged interactions to
define the intermolecular interactions in the Coulomb mimic system. Thus the
theory is not restricted to point counterions and very accurate results have already been found
for uniform fluids with charged hard cores~\cite{LMFCoulomb, LMFPairingAndWalls}.

However, additional approximations were made in these earlier applications
of LMF theory. In particular the Boltzmann approximation for the density response
to the effective field was used for the charged wall
system~\cite{LMFPairingAndWalls}.  Here we use Monte
Carlo (MC) simulations to accurately determine the density
response.  We believe that such simulations of the mimic system will often be needed to
obtain quantitative results from LMF theory in more realistic models of
biophysical interest.

Then the only remaining errors are those inherent in the LMF mapping itself.
The results provide a critical test of LMF theory in a nonuniform Coulomb
system where the basic physics of the counterion-mediated attraction is highly
nontrivial but well-understood, and where extensive benchmark simulations are
available~\cite{MoreiraNetz}.  

\begin{figure}[tbp]
  \centering
  \includegraphics*[width=3.375in]{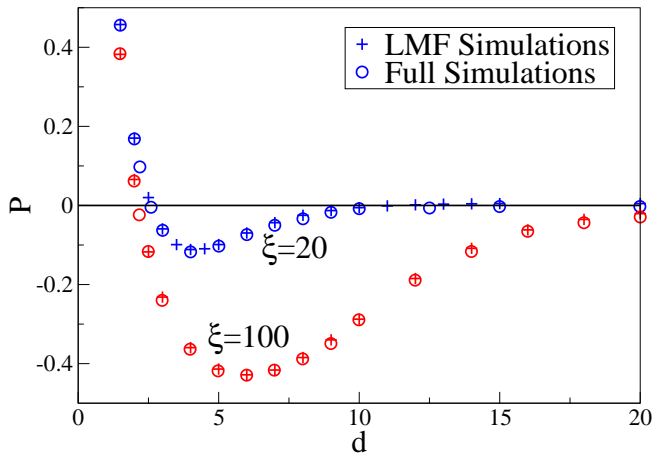}
  \caption{Pressure vs.~distance curves for two couplings,
    $\xi=20$ and $\xi=100$.  LMF simulations agree very well with
    full simulations from~\cite{MoreiraNetz}. For $d=20$
    and $\xi =20$, the effective core size $\sigma_{min}=18$ and for $d=20$
    and $\xi =100$, $\sigma_{min}=34$.}
  \label{fig:PvsD}
\end{figure}

Before giving details of the model and the LMF mapping, the quantitative
accuracy achieved in practice is illustrated in Fig.~\ref{fig:PvsD}, which compares the
dimensionless osmotic pressure $P$ for the full Coulomb
system~\cite{MoreiraNetz} to results of the LMF theory for two system
couplings, $\xi$.  Depending on the separation $d$, either coupling can result
in a net attractive force on the walls as indicated by negative values of $P$.
Simulations of the full Coulomb system required careful and costly treatment
of periodic boundary conditions using the Lekner-Sperb method; our simulations
of the short-ranged mimic system used only a simple minimum image method.

In the model two infinite hard walls with a negative charge density $q_w$ are
located at $z=0$ and $z=d$. Positive neutralizing point counterions with
valence $Z$ and charge $Ze_0$ are contained in $0 \leq z \leq d$ and there is
a uniform dielectric constant $\epsilon$ everywhere. The three length scales
shown in Fig.~\ref{fig:FullModel} can be motivated by an examination of the
energetics of \emph{one} uniformly charged wall at $z_w$.  The potential
energy between the wall and a counterion at $z$ is $-2\pi q_w Ze_0\left|z -
  z_w\right|/\epsilon $.  The distance where this potential equals $k_BT$
defines the \emph{Gouy-Chapman length} (including ion valence) $L_G \equiv
\epsilon k_B T/|2 \pi q_w Z e_0|$. The \emph{Bjerrum length} $L_B$ including ion
valence is similarly defined using the potential energy between a pair
of ions: $L_B \equiv Z^2 e_0^2/(\epsilon k_B T)$.  The third length scale,
$L_w$, is determined from the surface area of the wall neutralized by one
counterion: $L_w^2 \equiv Z e_0/\left|q_w\right|=2 \pi L_B L_G$.

We will use dimensionless variables where lengths are measured in units of
$L_G$ and energy in units of $k_BT$.  Specifying $d$ and the coupling strength
$\xi \equiv L_B / L_G$ fully defines the thermodynamic state of this system.
Effective attractions can arise for strong-coupling states with
$\xi \gtrsim 12$~\cite{MoreiraNetz}.  Since the
total force on a counterion from both walls exactly cancels,
the bare external potential $\phi=0$ for $0 \leq z \leq d$. 
However due to the long-ranged
Coulomb repulsion, counterions will organize next to the walls
into either one or two layers, based on a complex balance between
coupling strength $\xi$ and the width $d$ available.

Figure~\ref{fig:FullModel} qualitatively depicts a weakly attractive state
with $\xi=20$ and $d=10$. Here, most counterions are found in separate
two-dimensional (2D) liquid layers near each wall, with a characteristic
nearest neighbor spacing of order $L_w=11.2$ fixed by local
neutrality.  The effective attractions arise mainly from cross-correlations
between ions in the two layers, and become even stronger at smaller
separations when the counterions are forced into a \emph{single} 2D
layer with characteristic spacing $L_w/\sqrt{2}$~\cite{RouzinaBloomfield}.

Figure~\ref{fig:LMFModel} gives the LMF mapping to the mimic system.  Mimic
ions have a short-ranged repulsive Coulomb core with size
$\sigma_{min}$ of
order $L_w$ (indicated by solid circles), and their averaged long-ranged
repulsions lead to an effective external field $\phi_R(z)$ with wells of depth
$1.5k_B T$ at the walls.  At still larger $d$,
the wells deepen and mimic particles form two distinct layers. 
At smaller $d$, the wells disappear and the 
Coulomb cores from separate layers overlap significantly, forcing the system
into a single 2D layer. 

The derivation of LMF theory and its application to general Coulombic systems
are given in detail
elsewhere~\cite{WeeksYBG2,WeeksLMF,LMFCoulomb,LMFPairingAndWalls}.  The basic
ideas can most easily be seen for a simple one-component system in an external
field $\phi (\mathbf{r})$, where the intermolecular pair potential
$w(r)=u_0(r)+u_1(r)$ is properly separated into a short-ranged ``core'' part
$u_0(r)$ and a slowly-varying longer-ranged $u_1(r)$.  The mimic system is
composed of pair interactions $u_0(r)$ and a renormalized or effective
external field $\phi_R(\mathbf{r})$, which is supposed to induce a nonuniform
singlet density in the mimic system (indicated by the subscript R) equal to
that induced by $\phi$ in the full system:
\begin{equation}
  \rho_R(\mathbf{r};[\phi_R])=\rho(\mathbf{r};[\phi]).
\label{eqn:LMFstructure}
\end{equation}
This defines a mapping relating structure in the mimic and full systems.
The explicit LMF equation for $\phi_R(\mathbf{r})$
incorporates a density-weighted average
over the slowly-varying interactions $u_1$ and is given up to a constant by
\begin{equation}
  \phi _{R}(\mathbf{r})=\phi (\mathbf{r})+\int d{\mathbf{r}}^{\prime }\rho
  _{R}(\mathbf{r}^{\prime };[\phi_R])u_{1}(|\mathbf{r}^{\prime
  }-\mathbf{r}|). 
\label{eqn:LMFgeneral}
\end{equation}

This equation can be derived by integrating the first equation of the exact
Yvon-Born-Green hierarchy relating intermolecular forces and conditional
singlet densities in the full and mimic systems after making two
interconnected and physically reasonable
approximations~\cite{WeeksYBG2,WeeksLMF}.  First, when
Eq.~(\ref{eqn:LMFstructure}) holds, the conditional singlet density $\rho
_{R}({\bf r}^{\prime }|{\bf r};[\phi_R])$ in the mimic system should also
approximately equal that in the full system, provided that $u_0$ gives a good
representation of the short-ranged core interactions between particles.
Second, the force from the slowly-varying $u_1(|{\mathbf{r}}^{\prime
}-{\mathbf{r}}|)$ should be very small over the range $\bar{a}$ of characteristic
nearest neighbor distances where $\rho _{R}(\mathbf{r}^{\prime
}|\mathbf{r};[\phi_R])$ differs significantly from
$\rho_R(\mathbf{r}^{\prime};[\phi_R])$.  Then $\rho _{R}(\mathbf{r}^{\prime
}|\mathbf{r};[\phi_R])$ may be reasonably replaced by
$\rho_R(\mathbf{r}^{\prime};[\phi_R])$ in the integration of the force that
yields Eq.~(\ref{eqn:LMFgeneral}).

\begin{figure}[tbp]
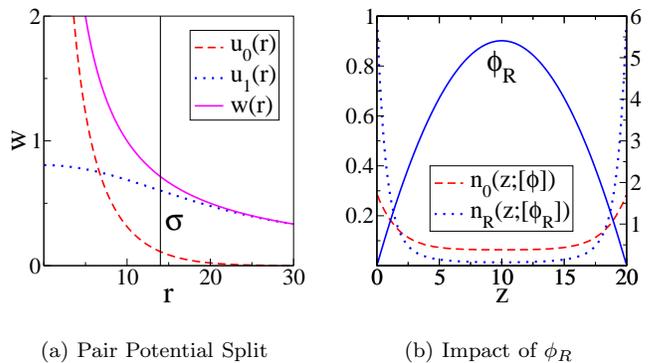

  \centering
  \subfigure[Pair Potential Split]{
    \label{fig:potsplit}
    \includegraphics*[height=1.55in]{potsplit.eps}}
  \hspace{0.075in}
  \subfigure[Impact of $\phi_R$]{
    \label{fig:impactphiR}
    \includegraphics*[height=1.55in]{ImpactPhiR.eps}}
  \caption{Potentials and densities for $\xi=10$ with $\sigma=14$.
    In~\subref{fig:potsplit} $\xi/r$ is split into $u_0$ and $u_1$.  As
    shown in \subref{fig:impactphiR} for $d=20$, when simulating
    mimic particles that interact only via $u_0$, the inclusion of $\phi_R$,
    which here has well depths greater
    than $5 k_B T$ (scale on right) rather than the flat $\phi$, is crucial.
    $n_0(z;[\phi])$ is quite different than the accurate $n_R(z;[\phi_R])$
    as shown by the contact densities (scale on left).}
  \label{fig:Potentials}
\end{figure}

For Coulomb systems we can control the accuracy of the second (mean field)
approximation by convoluting the dimensionless Coulomb potential $w(r)=\xi /r$
with a Gaussian whose width $\sigma$ is a parameter at our
disposal~\cite{LMFCoulomb,LMFPairingAndWalls}.  This yields a long-ranged
component $u_1(r)=\xi \erf(r/\sigma )/r$ that remains slowly varying at
distances less than $\sigma$ as illustrated in Fig.~\ref{fig:potsplit}, and
the associated core component $u_0(r)=\xi \erfc(r/\sigma )/r$. 
When $\sigma$ is too small,
results of the LMF theory will be poor and will
vary rapidly as $\sigma$ increases. But when $\sigma \geq \sigma_{min}$,
with $\sigma_{min}$ of order the characteristic neighbor
spacing $\bar{a}$, we expect that $u_1$ is sufficiently
slowly-varying that the LMF averaging is consistent and there will be little change
in results as $\sigma$ increases beyond $\sigma_{min}$~\cite{LMFCoulomb,LMFPairingAndWalls}.

Using this choice of
$u_1$, and noting that $\phi =0$, we can integrate
exactly over lateral coordinates in Eq.~(\ref{eqn:LMFgeneral}) and obtain the
two-wall LMF equation:
\begin{equation}
\phi_R (z) = \int_0^d \, dz^{\prime} n_R(z^{\prime};[\phi_R]) G(z^{\prime},z).
\label{eqn:LMFwall}
\end{equation}
Here $n_R(z) \equiv 2 \pi \xi \rho_R(z)$ is a dimensionless rescaled density
and $G(z^\prime,z) \equiv -|z - z^\prime|
\erf\left(|z-z^\prime|/{\sigma}\right) -\sigma \pi^{-1/2}\exp\left[-
  (z-z^\prime)^2/{\sigma}^2\right] + C(z^\prime)$ can be interpreted as the
potential at $z$ due to a Gaussian charge density $\sigma
\pi^{-1/2}\exp\left[-(z-z^\prime)^2/{\sigma}^2\right]$, with $C(z^\prime)$
chosen so that $G(z^\prime,0)=0$.  As explained earlier, $\phi_R(z)$ plays an
important role in this nonuniform mimic system.  The bare field
$\phi(z)=0$ must be replaced by $\phi_R(z)$ in simulations of the mimic system
for the particles to separate correctly into two layers, as
shown in Fig.~\ref{fig:impactphiR}. 

LMF theory gives exact results
both as $\xi \rightarrow 0$, where
it reduces to the Poisson-Boltzmann (PB) theory, and in the strong coupling limit
$\xi \rightarrow \infty$  \cite{LMFPairingAndWalls}.
To assess its performance for intermediate $\xi$, canonical MC simulations
of the nonuniform mimic system were carried out in a
simulation cell of volume $L\times L\times d$.  Particle number $N$ was chosen
so that the wall length $L$ dictated by neutrality is large enough to justify
the minimum image convention.  We used the simplest self-consistent simulation
closure of the LMF equation by explicitly iterating solutions indexed by $i$ of
Eq.~(\ref{eqn:LMFwall}) with converged MC values for $n_R(z)$ until the
self-consistency criterion $\int dz \left| \phi_R^{(i)} - \phi_R^{(i-1)}
\right|/d < 0.001$ was met. Properties were averaged over $5\times10^{5}$ --
$2\times 10^{6}$ simulation sweeps.  $n_R(z)$ and $\phi_R(z)$ were calculated on
a grid with spacing $\Delta z = \min \left\{0.1,0.01d\right\}$.
The reduced osmotic pressure was calculated by an accurate method that uses
the well-converged density at the midplane and the force between particles
and walls on the left and right of the
midplane~\cite{MoreiraNetz,ValleauMidplane,MOP}: $P =
n\left(z_{midplane}\right) + 2 \pi \xi \left<F_{LR}\right>/A $.

Crucial for the success of the theory is the proper choice of $\sigma_{min}$,
which should scale with the characteristic neighbor distance $\bar{a}$
at strong coupling \cite{LMFCoulomb,LMFPairingAndWalls}.
The discussion above suggests a simple criterion that uses the consistency of the
theory itself to determine a precise value for $\sigma_{min}$.  During a
simulation with a given $\sigma$, we measure the nearest-neigbor distance
averaged over particles, $\left<L_{nn}\right>$.  As $\sigma$ increases from
small values by steps $j$, we expect initially that $\left<L_{nn}\right>$ will
increase as the core repulsions in $u_0(r)$ becomes larger. But for $\sigma
\ge \sigma_{min}$, the variation in $\left<L_{nn}\right>$ should level off.
As a numerical criterion that gives very reasonable results we choose
the first $\sigma$ such that
$\left( \left<L_{nn}^j\right> - \left<L_{nn}^{j-1}\right> \right) /
\left(\sigma^j - \sigma^{j-1}\right) < 0.005$.

\begin{figure}[tbp]
  \centering
  \subfigure[$\sigma$ Convergence]{
    \label{fig:SigConv}
    \includegraphics*[height=1.5in]{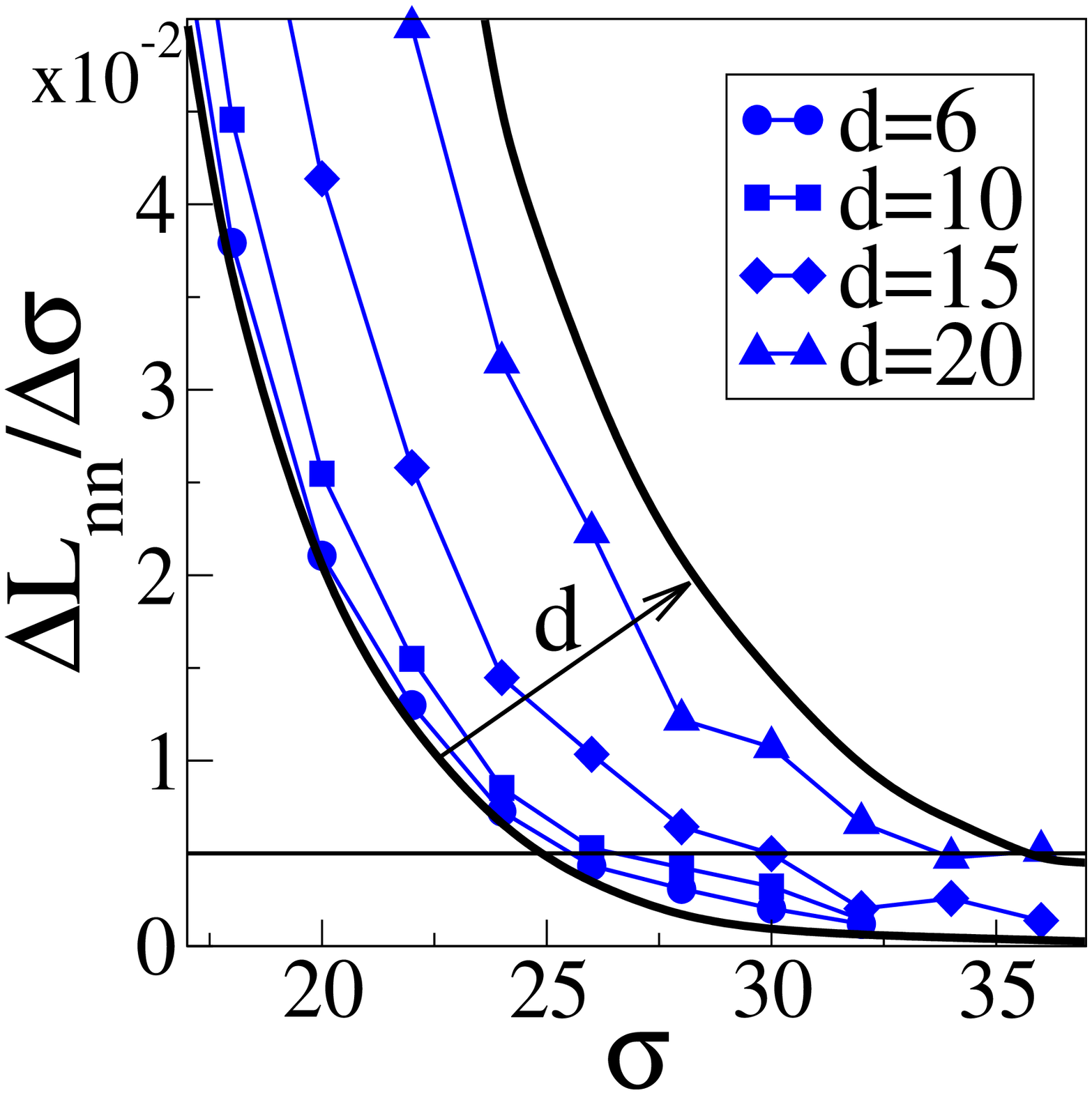}} 
  \hspace{0.075in}
  \subfigure[Comparison to Correlations]{
    \label{fig:GSame}
    \includegraphics*[height=1.5in]{GSame.eps}}
  \caption{Determination of  the \emph{consistency parameter} $\sigma_{min}$.
    The criterion is applied for $\xi=100$ and varying $d$
    in~\subref{fig:SigConv}; the horizontal line indicates the convergence
    threshold of 0.005.  The smooth curves schematically represent the upper and lower
    limits for $\sigma$ convergence; once a single layer or two layers
    unambiguously form, the neighbor distance $\bar{a}$
    and hence $\sigma_{min}$ should not vary further.
    $\left<L_{nn}\right>$ (dotted lines) and $\sigma_{min}$ (dashed lines) are
    related to $g_{2D}$ pair correlation curves in~\subref{fig:GSame}.}
  \label{fig:LnnSigCalc}
\end{figure}

Figure~\ref{fig:LnnSigCalc} illustrates the application of this convergence
criterion to a strongly-coupled system with $\xi=100$.  In
Fig.~\ref{fig:SigConv} the smoothed curves qualitatively represent the limits
of $\sigma_{min}$ determined for two separate layers at large $d$ and for a
single layer at small $d$.  Between those curves lie data for specific $d$,
ranging from two weakly correlated separate layers with weak attraction
($d=20$) to a single layer configuration where attraction is maximal ($d=6$).
As the walls are pushed closer together and the counterions shift from two
layers to one layer, the characteristic neighbor spacing $\bar{a}$ decreases
by a factor of $\sqrt{2}$.  In the $\sigma_{min}$ convergence plots, we see
the expected corresponding factor of $\sqrt{2}$ as $\sigma_{min}$ shifts from
36 to 26.

Figure~\ref{fig:GSame} shows how $\left<L_{nn}\right>$ and $\sigma_{min}$
relate to a special correlation function proposed by Rouzina and Bloomfield to
better probe the nature of correlations between particles in the 2D
layers~\cite{RouzinaBloomfield}. The $g_{2D}(r_{||})$ shown here is the
correlation function for pairs of particles on the same side of the midplane,
with distances projected onto the $xy$-plane.  Note that $\sigma_{min}$ is
larger than the distance $\bar{a}$ of
the first peak in $g_{2D}(r_{||})$, ensuring that $u_1$ varies slowly
over this nearest neighbor distance.

\begin{figure}[tbp]
  \centering
  \includegraphics*[width=3.375in]{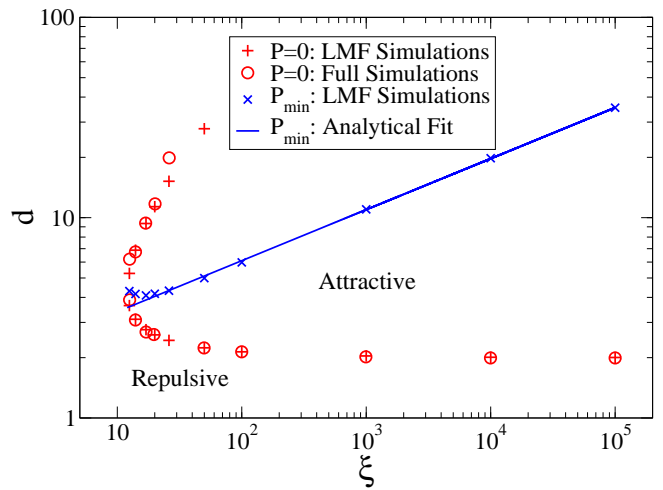}
  \caption{Points where $P(\xi,d)=0$ from
    LMF simulations are compared to full system simulations
    in~\cite{MoreiraNetz}.  In addition, separations $d_{min}$ where the
    maximum attractive force is found are fit by a power law to $L_W\propto
    \sqrt{\xi}$ ($d_{min} = 1.18 L_W^{0.509}$). This agrees very well with an
    LMF scaling prediction~\cite{LMFPairingAndWalls}.}
  \label{fig:Phase}
\end{figure}

The ``phase diagram'' of ($\xi$,$d$) points where $P=0$ is given in
Fig.~\ref{fig:Phase}.  LMF results again agree very well with those of the
full system. Also shown are points where LMF simulations yield the minimum
pressure $P_{min}$ (maximum attractive force).  As argued in
Ref.~\onlinecite{LMFPairingAndWalls}, $P_{min}$ should occur at a separation
$d_{min}$ where a single layer of counterions first forms.  By making a simple
approximation for $\phi_R$ and physically connecting $\sigma_{min}$ to
$\bar{a} \propto L_w$, we find $d_{min} \propto \sqrt{L_w}$.  The simulation
results for $d_{min}$ are best fit by $d_{min}=1.18 L_w^{0.509}$, which agrees
quite well with this scaling prediction~\cite{LMFPairingAndWalls}.

Neither the use of an effective field in LMF theory nor simulations with
short-ranged Coulomb cores is new.  Implicit in the classical
PB or Gouy-Chapman theory is an effective field resulting
from a mean-field average of the Coulomb interactions.  However, the PB
approach includes the rapidly varying Coulomb core components in the average
(effectively choosing $\sigma=0$) and is accurate only for dilute
weakly-coupled systems.  Spherical truncations of Coulomb interactions as
suggested by generalized reaction field methods have had some notable
successes in simulations of dense strongly-coupled uniform
systems~\cite{HummerRF}. But truncated interactions alone give poor results
for geometrically nonuniform systems like water between
walls~\cite{Spohr,WaterTruncation} or our two-wall system, as
illustrated in Fig.~\ref{fig:impactphiR}.

In contrast, LMF theory provides a general conceptual framework for nonuniform
Coulomb systems.  It determines a physical choice for the short-ranged Coulomb
cores and uses mean field theory in a consistent way
to generate an effective potential that accounts for the remaining long-ranged
interactions. The classical PB approach is greatly improved
by averaging only over the slowly-varying $u_1$
as dictated by LMF theory~\cite{LMFCoulomb}
and can even predict attraction for the two-wall system~\cite{LMFPairingAndWalls}.
We have shown here that simulations of the short-ranged cores in conjunction with the
$\phi_R$ given by LMF theory give quantitative agreement with simulations of
the full Coulomb system.  Further results for this system including
analysis of the 2D correlation functions
during the formation of a single or two layers,  the scaling of
$\sigma_{min}$, and more realistic descriptions of counterions and co-ions
will be reported elsewhere.  This work was supported by NSF
through grant CHE05-17818. JMR was supported by an NDSEG fellowship.


\begin{thebibliography}{15}
\expandafter\ifx\csname natexlab\endcsname\relax\def\natexlab#1{#1}\fi
\expandafter\ifx\csname bibnamefont\endcsname\relax
  \def\bibnamefont#1{#1}\fi
\expandafter\ifx\csname bibfnamefont\endcsname\relax
  \def\bibfnamefont#1{#1}\fi
\expandafter\ifx\csname citenamefont\endcsname\relax
  \def\citenamefont#1{#1}\fi
\expandafter\ifx\csname url\endcsname\relax
  \def\url#1{\texttt{#1}}\fi
\expandafter\ifx\csname urlprefix\endcsname\relax\def\urlprefix{URL }\fi
\providecommand{\bibinfo}[2]{#2}
\providecommand{\eprint}[2][]{\url{#2}}

\bibitem[{\citenamefont{Levin}(2002)}]{LevinReview}
\bibinfo{author}{\bibfnamefont{Y.}~\bibnamefont{Levin}}, \bibinfo{journal}{Rep.
  Prog. Phys.} \textbf{\bibinfo{volume}{65}}, \bibinfo{pages}{1577}
  (\bibinfo{year}{2002}).

\bibitem[{\citenamefont{Gelbart et~al.}(2000)\citenamefont{Gelbart, Bruinsma,
  Pincus, and Parsegian}}]{DNAElectrostatics}
\bibinfo{author}{\bibfnamefont{W.~M.} \bibnamefont{Gelbart}},
  \bibinfo{author}{\bibfnamefont{R.~F.} \bibnamefont{Bruinsma}},
  \bibinfo{author}{\bibfnamefont{P.~A.} \bibnamefont{Pincus}},
  \bibnamefont{and} \bibinfo{author}{\bibfnamefont{V.~A.}
  \bibnamefont{Parsegian}}, \bibinfo{journal}{Phys. Today}
  \textbf{\bibinfo{volume}{53}} (\bibinfo{year}{2000}).

\bibitem[{\citenamefont{Rouzina and Bloomfield}(1996)}]{RouzinaBloomfield}
\bibinfo{author}{\bibfnamefont{I.}~\bibnamefont{Rouzina}} \bibnamefont{and}
  \bibinfo{author}{\bibfnamefont{V.~A.} \bibnamefont{Bloomfield}},
  \bibinfo{journal}{J. Phys. Chem.} \textbf{\bibinfo{volume}{100}},
  \bibinfo{pages}{9977} (\bibinfo{year}{1996}).

\bibitem[{\citenamefont{Boroudjerdi et~al.}(2005)\citenamefont{Boroudjerdi,
  Kim, Naji, Netz, Schlagberger, and Serr}}]{NetzReview2005}
\bibinfo{author}{\bibfnamefont{H.}~\bibnamefont{Boroudjerdi}},
  \bibinfo{author}{\bibfnamefont{Y.-W.} \bibnamefont{Kim}},
  \bibinfo{author}{\bibfnamefont{A.}~\bibnamefont{Naji}},
  \bibinfo{author}{\bibfnamefont{R.}~\bibnamefont{Netz}},
  \bibinfo{author}{\bibfnamefont{X.}~\bibnamefont{Schlagberger}},
  \bibnamefont{and} \bibinfo{author}{\bibfnamefont{A.}~\bibnamefont{Serr}},
  \bibinfo{journal}{Phys. Rep.} \textbf{\bibinfo{volume}{416}},
  \bibinfo{pages}{129} (\bibinfo{year}{2005}).

\bibitem[{\citenamefont{Katsov and Weeks}(2001)}]{WeeksLMF}
\bibinfo{author}{\bibfnamefont{K.}~\bibnamefont{Katsov}} \bibnamefont{and}
  \bibinfo{author}{\bibfnamefont{J.~D.} \bibnamefont{Weeks}},
  \bibinfo{journal}{J. Phys. Chem. B} \textbf{\bibinfo{volume}{105}},
  \bibinfo{pages}{6738} (\bibinfo{year}{2001}).

\bibitem[{\citenamefont{Weeks et~al.}(1998)\citenamefont{Weeks, Katsov, and
  Vollmayr}}]{WeeksYBG2}
\bibinfo{author}{\bibfnamefont{J.~D.} \bibnamefont{Weeks}},
  \bibinfo{author}{\bibfnamefont{K.}~\bibnamefont{Katsov}}, \bibnamefont{and}
  \bibinfo{author}{\bibfnamefont{K.}~\bibnamefont{Vollmayr}},
  \bibinfo{journal}{Phys. Rev. Lett.} \textbf{\bibinfo{volume}{81}},
  \bibinfo{pages}{4400} (\bibinfo{year}{1998}).

\bibitem[{\citenamefont{Chen et~al.}(2004)\citenamefont{Chen, Kaur, and
  Weeks}}]{LMFCoulomb}
\bibinfo{author}{\bibfnamefont{Y.-G.} \bibnamefont{Chen}},
  \bibinfo{author}{\bibfnamefont{C.}~\bibnamefont{Kaur}}, \bibnamefont{and}
  \bibinfo{author}{\bibfnamefont{J.~D.} \bibnamefont{Weeks}},
  \bibinfo{journal}{J. Phys. Chem. B} \textbf{\bibinfo{volume}{108}},
  \bibinfo{pages}{19874} (\bibinfo{year}{2004}).

\bibitem[{\citenamefont{Chen and Weeks}(2006)}]{LMFPairingAndWalls}
\bibinfo{author}{\bibfnamefont{Y.-G.} \bibnamefont{Chen}} \bibnamefont{and}
  \bibinfo{author}{\bibfnamefont{J.~D.} \bibnamefont{Weeks}},
  \bibinfo{journal}{Proc. Nat. Acad. Sci. USA} \textbf{\bibinfo{volume}{103}},
  \bibinfo{pages}{7560} (\bibinfo{year}{2006}).

\bibitem[{\citenamefont{Moreira and Netz}(2002)}]{MoreiraNetz}
\bibinfo{author}{\bibfnamefont{A.~G.} \bibnamefont{Moreira}} \bibnamefont{and}
  \bibinfo{author}{\bibfnamefont{R.~R.} \bibnamefont{Netz}},
  \bibinfo{journal}{Eur. Phys. J. E} \textbf{\bibinfo{volume}{8}},
  \bibinfo{pages}{33} (\bibinfo{year}{2002}).

\bibitem[{\citenamefont{Valleau et~al.}(1991)\citenamefont{Valleau, Ivkov, and
  Torrie}}]{ValleauMidplane}
\bibinfo{author}{\bibfnamefont{J.}~\bibnamefont{Valleau}},
  \bibinfo{author}{\bibfnamefont{R.}~\bibnamefont{Ivkov}}, \bibnamefont{and}
  \bibinfo{author}{\bibfnamefont{G.}~\bibnamefont{Torrie}},
  \bibinfo{journal}{J. Chem. Phys} \textbf{\bibinfo{volume}{95}},
  \bibinfo{pages}{520} (\bibinfo{year}{1991}).

\bibitem[{\citenamefont{Todd et~al.}(1995)\citenamefont{Todd, Evans, and
  Daivis}}]{MOP}
\bibinfo{author}{\bibfnamefont{B.}~\bibnamefont{Todd}},
  \bibinfo{author}{\bibfnamefont{D.~J.} \bibnamefont{Evans}}, \bibnamefont{and}
  \bibinfo{author}{\bibfnamefont{P.~J.} \bibnamefont{Daivis}},
  \bibinfo{journal}{Phys. Rev. E} \textbf{\bibinfo{volume}{52}},
  \bibinfo{pages}{1627} (\bibinfo{year}{1995}).

\bibitem[{\citenamefont{Hummer et~al.}(1996)\citenamefont{Hummer, Pratt, and
  Garcia}}]{HummerRF}
\bibinfo{author}{\bibfnamefont{G.}~\bibnamefont{Hummer}},
  \bibinfo{author}{\bibfnamefont{L.~R.} \bibnamefont{Pratt}}, \bibnamefont{and}
  \bibinfo{author}{\bibfnamefont{A.~E.} \bibnamefont{Garcia}},
  \bibinfo{journal}{J. Phys. Chem.} \textbf{\bibinfo{volume}{100}},
  \bibinfo{pages}{1206} (\bibinfo{year}{1996}).

\bibitem[{\citenamefont{Spohr}(1997)}]{Spohr}
\bibinfo{author}{\bibfnamefont{E.}~\bibnamefont{Spohr}}, \bibinfo{journal}{J.
  Chem. Phys} \textbf{\bibinfo{volume}{107}}, \bibinfo{pages}{6342}
  (\bibinfo{year}{1997}).

\bibitem[{\citenamefont{Feller et~al.}(1996)\citenamefont{Feller, Pastor,
  Rojnuckarin, Bogusz, and Brooks}}]{WaterTruncation}
\bibinfo{author}{\bibfnamefont{S.~E.} \bibnamefont{Feller}},
  \bibinfo{author}{\bibfnamefont{R.~W.} \bibnamefont{Pastor}},
  \bibinfo{author}{\bibfnamefont{A.}~\bibnamefont{Rojnuckarin}},
  \bibinfo{author}{\bibfnamefont{S.}~\bibnamefont{Bogusz}}, \bibnamefont{and}
  \bibinfo{author}{\bibfnamefont{B.~R.} \bibnamefont{Brooks}},
  \bibinfo{journal}{J. Phys. Chem.} \textbf{\bibinfo{volume}{100}},
  \bibinfo{pages}{17011} (\bibinfo{year}{1996}).

\end{thebibliography}
\end{document}